\documentclass[12pt]{article}
\pdfoutput=1
\usepackage{amssymb,amsmath}
\usepackage{amsbsy}
\usepackage{amsfonts}
\usepackage[dvips]{graphicx}
\usepackage{cancel}
\usepackage{enumerate}
\usepackage[colorlinks=true]{hyperref}
\newcommand{\AddrSanto}{
  {\it Universidad Santo Tomas, Colombia}}
\newcommand{\AddrUNAL}{
  {\it Departamento de F\'{\i}sica, Universidad Nacional de Colombia, Bogot\'a, Colombia}}

\title{Lepton Flavor Violation processes in 331 Models}

\author{J. M. Cabarcas$^1$\footnote{josecabarcas@usantotomas.edu.co}, J. Duarte$^2$\footnote{jduartec@unal.edu.co},
J.-Alexis Rodriguez$^2$\footnote{jarodriguezl@unal.edu.co}\\
1.\AddrSanto \\ 2.\AddrUNAL}
\date{}
\begin{document}
\maketitle{}

\abstract{Models based on the extended symmetry gauge $SU(3)_c \otimes SU(3)_L \otimes U(1)_Y$
can be build up with a leptonic sector consistent of five triplets in different $SU(3)_L$ representations where
additional heavy fermions are included. Some of these models present flavor changing neutral currents in the leptonic
sector which are mediated through the $Z'$ boson. One of these models is studied using the measurements of lepton flavor violation processes such as  $\tau\rightarrow lll$ with $l= e,\mu$ and $\mu(\tau) \to e (\mu) \gamma$. }


\section{Introduction}

In the framework of the standard model (SM) of high energy physics there are many issues unclear
that definitely requires extensions of the theory in the local symmetry and in the spectrum. One possible
alternative is based on the gauge symmetry  $SU(3)_c\otimes SU(3)_L\otimes U(1)_X$ known as 331 models\cite{331}. These
models can explain why there are three fermionic families and it is related with the number of colors in QCD through
the chiral anomaly cancellation condition. On the other hand, the models based on 331 symmetry are build in such a way that the couplings of the fermions
 with the new neutral $Z'$ boson are not universal in the interaction basis therefore
 in the mass eigenstates basis those couplings are not diagonal and flavor changing neutral currents (FCNC) at tree level arise up\cite{FCNC}. This is a special feature of the 331 models and it is because
one quark family is in a different representation of the gauge group to the other two families in order to satisfy the chiral anomaly cancellation condition. It is worth to mention that in some
331 models there are not only contribution of the left handed neutral current but also from the right handed neutral current.
 There are many studies of these new FCNC in the quark sector but there are not too many in the leptonic sector where
leptonic flavor violation  (LFV) processes at tree level are present.

In particular, LFV processes such as $\tau\rightarrow l^-l^+l^-$ with $l=e,\mu$ have been discussed in the framework
of the minimal supersimmetric standard model, Little Higgs models, left-right symmetry models and many other extensions
of the SM have been considered \cite{lfvmodels}. Some of these models predict branching fractions for $\tau\rightarrow l^-l^+l^-$ of the order of $10^{-7}$ which are in the range of possible detection of future experiments. Recently, MEGA and SINDRUM collaborations have reported new bounds on LFV processes, MEGA has reported $BR(\mu \rightarrow e\gamma) < 1.2 \times 10^{-11}$ and SINDRUM $BR(\mu \rightarrow 3e) <10^{-12}$ \cite{mega,sindrum}. These bounds together with the bounds on   $\tau\rightarrow lll$ with $l= e,\mu$ and $\mu(\tau) \to e (\mu) \gamma$ coming from BELLE and BABAR experiments are phenomenological sources to explore the origin of the mixing in the leptonic sector.

In general, the 331 models are classified depending how they cancel the chiral anomalies: there are two models that cancel out the anomalies requiring just one family and eight models where the three families are required. In the three family models, there are four models where the
leptons are treated identically, two of them treat two quark generations identically and finally, there are two models where all the lepton generations are treated differently  \cite{331}, here models without exotic charges will be considered. There is one of this 331 model where the leptonic
sector is described by five left handed leptonic triplets in different representations of the $SU(3)_L$ gauge group. Using these five leptonic representations is possible to get models where the
known three leptons coupled to the $Z'$ boson in very different respect to the new ones. This model is our interest in this work in order to study the LFV processes and therefore get some
constraints on the leptonic mixing matrix. In the next section we are going to show the main features of the model under consideration and then we focus on the LFV processes.

\section{The 331 Model}

The model considered below is based on the local gauge symmetry $SU(3)_C\otimes SU(3)_L \otimes U(1)_X$ (331),
where it is usual to write the electric charge generator as a linear
combination of the diagonal generators of the group as
 \begin{equation}
Q\ = \ T_3\,+\,\beta\, T_8\,+\,X\ .
\end{equation}
where the parameter $\beta$ is used to label the particular 331 model considered. For constructing the model
we choose $\beta= -1/\sqrt{3}$, which corresponds to models where the new fields in the spectra do not have exotic electric charges. The quark content of the model proposed is described by
\begin{eqnarray}
q_{mL} = \begin{pmatrix}
u_m \cr -d_m \cr B_m\end{pmatrix}_L\, \sim
(3^*,3,0),&\qquad& q_{3L}=\begin{pmatrix}
u_3 \cr d_3 \cr T_3
\end{pmatrix}_L\,\sim(3,3,1/3)\nonumber\\
d^c\sim (3^*,1,1/3),\qquad u^c\sim (3^*,1,-2/3),&\qquad& B_m^c\sim (3^*,1,1/3),\qquad T^c\sim (3^*,1,-2/3).
\end{eqnarray}
where $m=1,2$ and their quantum numbers under the 331 group are shown in the parenthesis. For the leptonic spectrum we use
\begin{eqnarray}
\Psi_{nL}=\begin{pmatrix}e_n^-\cr \nu_n \cr N_n^0
\end{pmatrix}_L\,\sim (1,3^*,-1/3)
,&\qquad& \Psi_{L}=\begin{pmatrix}
\nu_1 \cr e_1^-\cr E_1^-
\end{pmatrix}_L\,\sim (1,3,-2/3),\nonumber\\
\Psi_{4L}=\begin{pmatrix}
E_2^-\cr N_3^0\cr N_4^0
\end{pmatrix}_L\sim (1,3^*,-1/3), &\qquad& \Psi_{5L}=\begin{pmatrix}
N_5^0\cr E_2^+\cr e_3^+
\end{pmatrix}_L\sim (1,3^*,2/3),\nonumber\\
e_n^c\sim (1,1,1),\qquad e_3^c\sim (1,1,1),&\qquad& E_1^c\sim (1,1,1),\qquad E_2^c\sim (1,1,1).
\end{eqnarray}
\noindent with $n=2,3$. Five leptonic triplets plus the quark content are enough to insures cancellation of chiral anomalies. Notice that with this proposed assemble for the leptonic sector,
there is only one of the triplets that is not written in the adjoint representation of $SU(3)_L$
and it contains one of the standard lepton families of the SM.

On the other hand, in 331 models without exotic charges, the gauge bosons of the $SU(3)_L$ will
transform according to the adjoint representation  and the gauge boson field
$B_\mu$ is associated with the  $U(1)_X$ group which is a singlet under  $SU(3)_L$ and it does not have electric charge.
Once the gauge boson sector is identified then the neutral sector  $W^3$,
$W^8$ and $B$ is rotated to get the new neutral gauge bosons $A$, $Z$ and $Z'$, and they are
\begin{eqnarray}
\left(\begin{array}{c}
A \\ Z \\ Z^\prime
\end{array}\right) =
\left( \begin{array}{ccc}
S_W & - S_W/\sqrt{3} & C_W\sqrt{1- T_W^2/3} \\
C_W &  S_W T_W/\sqrt{3} & -\,S_W\sqrt{1- T_W^2/3} \\
0 & -\sqrt{1- T_W^2/3} &  -T_W/\sqrt{3}
\end{array} \right)
\left( \begin{array}{c}
W^3 \\ W^8 \\ B
\end{array} \right) \ ,
\label{bosgauneu}
\end{eqnarray}
where $\theta_W$ ($S_W = \sin\theta_W$, $C_W = \cos\theta_W$) is the Weinberg's angle defined by $T_W = \tan\theta_W
= g'/\sqrt{g^2+{g'}^2/3}$, with $g$ and $g'$ the coupling constants of the
$SU(3)_L$ and $U(1)_X$ groups respectively.
In this new basis, the photon  $A_\mu$ is the gauge boson associated to the charge  generator $Q$ while the $Z_\mu$ boson can be identified as the usual $Z$ gauge boson of the SM.

Our aim in this work concerns to the leptonic phenomenology, therefore only the leptonic sector will be addressed.
The Lagrangian for the neutral currents in this sector is
\begin{eqnarray}\label{lagrangiano}
{\cal L}_{NC}&=&-\sum_\ell\left[g S_W\, A_\mu\left\{\bar{\ell^0}\gamma_\mu\epsilon^{A}_{\ell_{(L)}}P_L{\ell}^0
+\bar{\ell^0}\gamma_\mu\epsilon^{A}_{\ell_{(R)}}P_R{\ell}^0 \right\}\right.\\\nonumber
&&\hskip1.cm+\frac{gZ^\mu}{2C_W}\left\{\bar{\ell^0}\gamma_\mu\epsilon^{Z}_{\ell_{(L)}}P_L{\ell}^0
+\bar{\ell^0}\gamma_\mu\epsilon^{Z}_{\ell_{(R)}}P_R{\ell}^0 \right\}\nonumber\\
&&\hskip1.cm+\left.\frac{g^\prime Z^{\prime\mu}}{2\sqrt{3}S_W C_W}\left\{\bar{\psi^0}\gamma_\mu\epsilon^{Z'}_{\ell_{(L)}}P_L{\ell}^0
+\bar{\ell^0}\gamma_\mu\epsilon^{Z'}_{\ell_{(R)}}P_R{\ell}^0 \right\}\right]
\end{eqnarray}
where $\ell^0$ in this notation stands for the charged leptons vector ${ \ell^{0\,T}}=\left(e_1^{0-},  e_2^{0-},  e_3^{0-}, E_1^{0-},  E_2^{0-}\right)$, the zero superscript
 denotes that the fields are in the interaction basis. The couplings to the neutral bosons are in the following matrices
\begin{eqnarray}\label{acople}
\epsilon_{ { \ell}_{L}}^{A} &=& I_{5\times5} \, ,\nonumber\\
\epsilon_{{\ell}_{(R)}}^{A}&=& I_{5\times5}\, , \nonumber \\
\epsilon_{{ \ell}_{L}}^{\it Z} &=& Diag(C_{2W},C_{2W},C_{2W},-2S_W^2,C_{2W})\, ,\nonumber\\
\epsilon_{{ \ell}_{R}}^{\it Z} &=& Diag(-2S_W^2,-2S_W^2,-2S_W^2,-2S_W^2,C_{2W})\, ,\nonumber\\
\epsilon_{{ \ell}_{L}}^{\it Z'} &=&  Diag(1,-C_{2W},-C_{2W},-C_{2W},-C_{2W})\, , \nonumber\\
\epsilon_{{ \ell}_{R}}^{\it Z'} &=&  Diag(2S_W^2,2S_W^2,-C_{2W},2S_W^2,1)\, ,
\end{eqnarray}
 where $C_{2W}=\cos{(2\theta_W)}$. The couplings of the charged leptons to the photon
$A_\mu$ are universal as well as the couplings of the standard leptons to the $Z$ boson. A feature of this model is that the couplings
of the standard left handed leptons as well as the right handed leptons to the $Z'$ boson are not universal,
 due to the fact that one of the lepton triplets is in a different representation than the other two. Since the couplings of the $Z'$ boson to the standard leptons are not universal then the obtained mixing matrix  will allow LFV at tree level when they are rotated to mass eigenstates.

 A similar procedure in the neutral leptonic sector can be done, we use the vector
 ${ N^{0\,T}}=\left(\nu_1^0,  \nu_2^0,  \nu_3^0,  N_1^0,  N_2^0,  N_3^0,  N_4^0, N_5^0\right)$ generating the couplings
\begin{eqnarray}
\epsilon_{{ N}_{L}}^{A}&=&0  \, \nonumber\\
\epsilon_{{ N}_{L}}^{Z} &=& Diag(1,1,1,0,0,1,0,-1)\,  \nonumber \\
\epsilon_{{ N}_{L}}^{Z'} &=& Diag(1,-C_{2W},-C_{2W},2C_W^2,2C_W^2,-C_{2W},2C_W^2,-1) .
\end{eqnarray}
Therefore, the couplings of the standard neutrinos to the photon $A$ and $Z$ boson are universal but the couplings of these leptons to the $Z'$ are not.\\

It is possible to re-write the neutral current Lagrangian in order to
use the formalism presented in reference \cite{Langacker:2000ju} and generate an effective Lagrangian like
\begin{equation}
{\cal L}_{\rm NC}^{eff} \ = \ - \; e\, J_{em}^\mu\,A_\mu\; -
\; g_1\,J^{(1)\mu}\,Z_{1\mu}\; - \; g_2\,J^{(2)\mu}\,Z_{2\mu}  \ ,
\end{equation}
where the currents associated to the gauge  $Z$ and $Z'$ bosons are
\begin{eqnarray}
J^{(1)}_\mu & = & \sum_{ij} \bar \ell_i^0\, \gamma_\mu\, (\epsilon^Z_{\ell_{L}}
\, P_L + \epsilon^Z_{\ell_{R}} \, P_R) \ell_j^0 \ \ , \\
J^{(2)}_\mu & = & \sum_{ij} \bar \ell_i^0\, \gamma_\mu\, (\epsilon^{Z'}_{\ell_{L}}
\, P_L + \epsilon^{Z'}_{\ell_{R}} \, P_R) \ell_j^0\ \ ,
\end{eqnarray}
with $g_1 = g/C_W$. The $\ell_i^0$ leptons and the gauge bosons  $Z_1$ and $Z_2$ are interaction eigenstates and
 the matrices  $\epsilon^{Z}_{\ell_{L,R}}$ and $\epsilon^{Z'}_{\ell_{L,R}}$ in the charged sector were defined
in equation ~\eqref{acople}. When the fields of the theory are rotated to mass or physical eigenstates the effective Lagrangian for the charged leptons can be finally written as:
\begin{equation}
{\cal L}_{\rm eff} \ = \ -\;\frac{4\,G_F}{\sqrt{2}} \sum_{ijkl} \sum_{XY}
\; C^{ijkl}_{XY}\ (\overline{\ell}_i\,\gamma^\mu\, P_X\,\ell_j)\
(\overline{\ell}_k\,\gamma_\mu\, P_Y\,\ell_l)\ ,
\end{equation}
where $X$ and $Y$ run over the chiralities $L,R$ and indices $i,j,k,l$ over the leptonic families. The coefficients $C^{ijkl}_{XY}$ for the stantard leptons, assuming a mixing angle  $\theta$
between $Z$ and $Z'$ bosons,  are given by \cite{Langacker:2000ju},
\begin{equation}
C^{ijkl}_{XY} \ =
z\,\rho\,\bigg(\frac{g_2}{g_1}\bigg)^2 \,B^{X}_{ij}\, B^{Y}_{kl}\ ,
\label{cijkl}
\end{equation}
where
\begin{eqnarray}
\rho \ & = & \ \frac{m_W^2}{m_Z^2 C^2_W}\,, \nonumber \\
z & = & \left(\sin^2\theta + \frac{m_Z^2}{m_{Z'}^2} \cos^2\theta\right)\nonumber \,,\\
\left(\frac{g_2}{g_1}\right)^2 &= & \frac{1}{3(1-4S_W^2)} \, .
\label{zz}
\end{eqnarray}
The  $B^X$ corresponds to the
 matrices obtained when the unitary matrices  $V^\ell_{L,R}$ are introduce to get the mass eigenstates and to diagonalize the Yukawa coupling matrices, particularly
\begin{equation}
{B^X}_{} \ = \ {V_X^{\ell\,\dagger}\;\epsilon_\ell^{Z'}\; V_X^\ell}_{}\ .
\label{bes}
\end{equation}

For the matrix $V$ we will use a well accepted Ans\"{a}tz \cite{FCNC} where
\begin{equation}
V_L^\ell \ = \ P \; \tilde V\; K
\label{vdef}
\end{equation}

with $P = {\rm diag}(e^{i\phi_1},1,e^{i\phi_3})$, $K={\rm
diag}(e^{i\alpha_1},e^{i\alpha_2},e^{i\alpha_3})$, and the unitary matrix  $\tilde V$ can be parameterized
using three standard mixing angles $\theta_{12}$, $\theta_{23}$ and $\theta_{13}$ and a phase $\varphi$,
\begin{equation}
\tilde V \ = \left(
\begin{array}{ccc}
c_{12}\,c_{13} & s_{12}\,c_{13} & s_{13}\,e^{-i\varphi} \\
-s_{12}\,c_{23}-c_{12}\,s_{23}\,s_{13}\,e^{i\varphi} &
c_{12}\,c_{23}-s_{12}\,s_{23}\,s_{13}\,e^{i\varphi} &
s_{23}\,c_{13} \\
s_{12}\,s_{23}-c_{12}\,c_{23}\,s_{13}\,e^{i\varphi} &
-c_{12}\,s_{23}-s_{12}\,c_{23}\,s_{13}\,e^{i\varphi} &
c_{23}\,c_{13}
\end{array}
\right)\ \ .
\label{vparam}
\end{equation}
Notice that if we are considering only the standard charged leptons the coupling matrices ec.(\ref{acople}) might be written as
\begin{eqnarray}
\label{eps2}
\epsilon_{{ \ell}_{L}}^{\it Z'} &=&  -(1-2S_W^2) {\bf I}_{3\times 3}+2\,C_W^2\, Diag(1,0,0)\, , \nonumber\\
\epsilon_{{ \ell}_{R}}^{\it Z'} &=&  2\,S_W^2\,{\bf I}_{3\times 3}-Diag(0,0,1)\, ,
\end{eqnarray}
At this point, we have to pointed out that the terms which are proportionals to the identity are not contribuiting
to the LFV processes at tree level while the second term in the above equations do. These equations (\ref{eps2}) correspond to the case when the first family is in the adjoint representation however if the second family was the chosen one to be in a different representation then the only change is in the second term which is proportional to $Diag(0,1,0)$ and if instead of that, the third family was the chosen one then again the only change is the position of the number one in the second term. We should emphasize that the source of LFV in neutral currents mediated by the $Z'$ boson, arise up from the non-diagonal elements
in the $3\times 3$ matrices $B^\ell_{L,R}$.

\section{LFV processes}

Our next task is to get bounds on the parameters involved in the LFV couplings and it is done considering different LFV processes.
Recently, the BABAR\cite{datosexperimentalesbabar}
and BELLE \cite{datosexperimentalesbelle} collaborations have reported measurements of various
LFV channels and they have put new bounds on these branching fractions, see table \ref{tab1}. Other channels to consider are $BR(\mu^-\rightarrow e^-\gamma)< 2,4 \times 10^{-12}$ and
$BR(\mu^-\rightarrow e^-e^-e^+)< 1,0 \times 10^{-12}$.

\begin{table}
 \centering
\begin{tabular}{|c|c|c|c|}
\hline\hline
Processes & $BR(\times 10^{-8})$ \cite{datosexperimentalesbelle} & $BR(\times 10^{-8})$ \cite{datosexperimentalesbabar} \\\hline
$\tau^-\rightarrow e^-\gamma$ & 12 & 3.3 \\\hline
$\tau^-\rightarrow \mu^-\gamma$ & 4.5 & 4.4 \\\hline
$\tau^-\rightarrow e^-e^+e^-$ & 2,7 & 2,9 \\\hline
$\tau^-\rightarrow \mu^-\mu^+\mu^-$ & 2,1 & 3,3 \\\hline
$\tau^-\rightarrow e^-\mu^+\mu^-$ & 2,7 & 3,2\\\hline
$\tau^-\rightarrow\mu^-e^+e^-$ & 1,8 & 2,2 \\\hline
$\tau^-\rightarrow e^+\mu^-\mu^-$ & 1,7 & 2,6\\\hline
$\tau^-\rightarrow \mu^+e^-e^-$ & 1,5 & 1,8 \\\hline
\end{tabular}
\caption{Experimental data and their bounds from BELLE (column 2) and BABAR (column 3)}
\label{tab1}
\end{table}

In the framework of the 331 model that we have already presented in section 2, we calculate the decay widths for the different processes that we are going to take into account. For
the $l_j\rightarrow l_i\gamma$ processes, the decay widths are

\begin{eqnarray}
\Gamma(l_j\rightarrow l_i\gamma) = && \frac{\alpha G_F^2 M_j^3}{8\pi^4} \left(\frac{g_2}{g_1}\right)^4\rho^2
 \left[\left(B^R M_l B^L\right)^2_{ij}+ \left(B^L M_l B^R\right)^2_{ij}  \right]\nonumber\\
\end{eqnarray}
with $i,j=e,\mu,\tau$, and $M_l$ a diagonal mass matrix where the electon mass has been neglected.

\begin{figure}[htp]
 \begin{center}
\includegraphics[width=6cm]{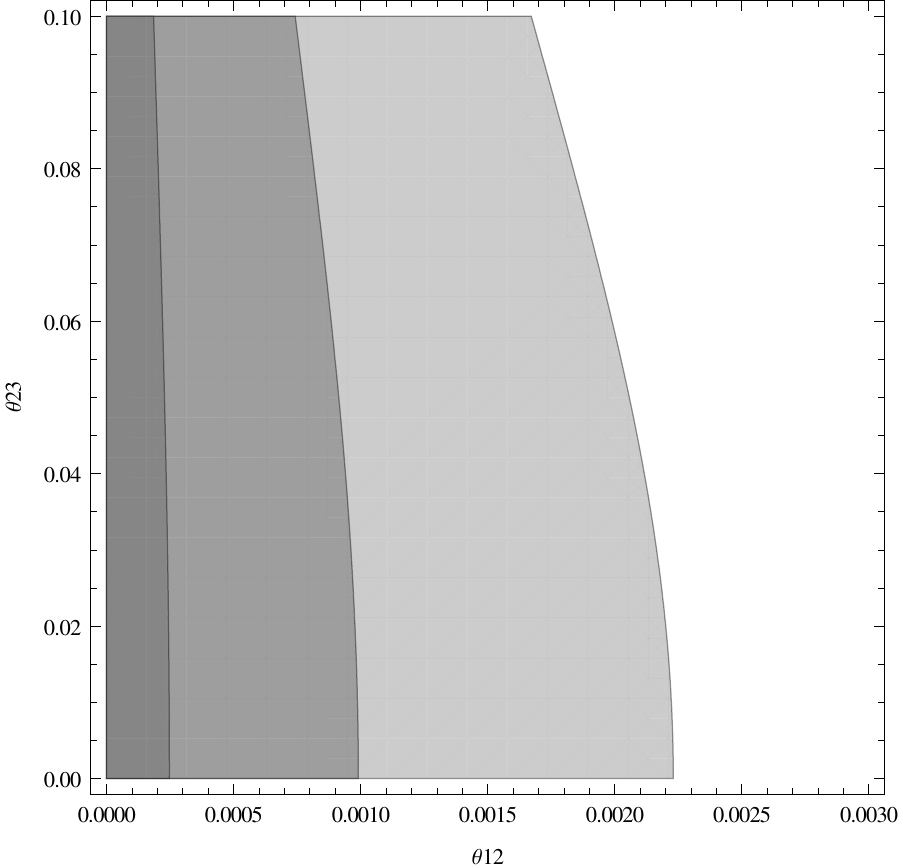} \quad
\includegraphics[width=6cm]{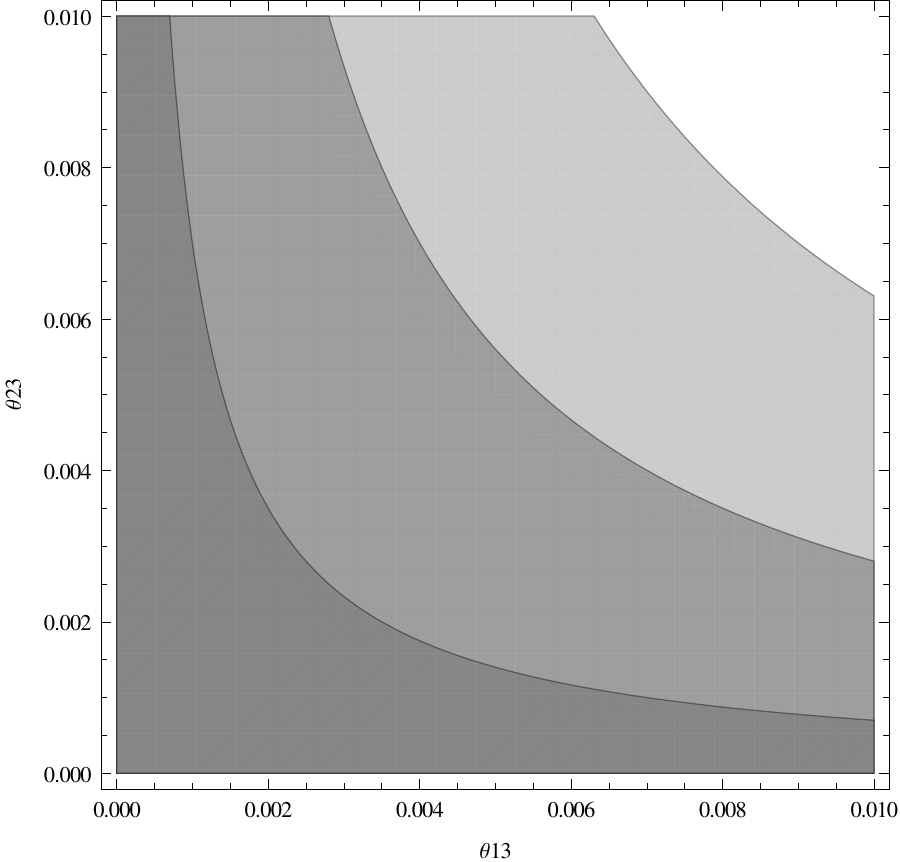}
\end{center}
\caption{Bounds coming from the $\mu\to e\gamma$ process in the different planes such that the third mixing angle is setting to zero for $m_{Z'}(1000, 2000, 3000 GeV)$.}
\label{1muegamma}
\end{figure}

From the table 1, we must evaluate the decay widths into three charged leptons, they are
\begin{eqnarray}
\Gamma(l_j\to l_i^- l_i^- l_i^+ )= &&\frac{G_F^2 M_{l_j}^5}{48 \pi^3}\left(\frac{g_2}{g_1}\right)^4 \rho^2\nonumber\\
&\times&\left[2 \left|B_{ij}^L B_{ii}^L\right|^2+2 \left|B_{ij}^R B_{ii}^R\right|^2+ \left|B_{ij}^L B_{ii}^R\right|^2+
\left|B_{ij}^R B_{ii}^L\right|^2\right] \, ,\nonumber\\
\Gamma(l_j\to l_i^- l_k^- l_l^+ )= &&\frac{G_F^2 M_{l_j}^5}{48 \pi^3}\left(\frac{g_2}{g_1}\right)^4 \rho^2\nonumber\\
&\times&\left[ \left|B_{ij}^L B_{kl}^L+B_{kj}^L B_{il}^L\right|^2+ \left|B_{ij}^R B_{kl}^R+B_{kj}^R B_{il}^R\right|^2+
\left|B_{ij}^L B_{kl}^R\right|^2+\left|B_{kj}^L B_{il}^R\right|^2 \right.\nonumber\\
&& +\left.
\left|B_{ij}^R B_{kl}^L\right|^2+\left|B_{kj}^R B_{il}^L\right|^2\right]\nonumber\\
\end{eqnarray}
where the elements $B_{ij}^{L,R}$ are defined in equation (~\ref{bes}) and $\rho$ in equation (~\ref{zz}).

\begin{figure}[htp]
 \begin{center}
\includegraphics[width=6cm]{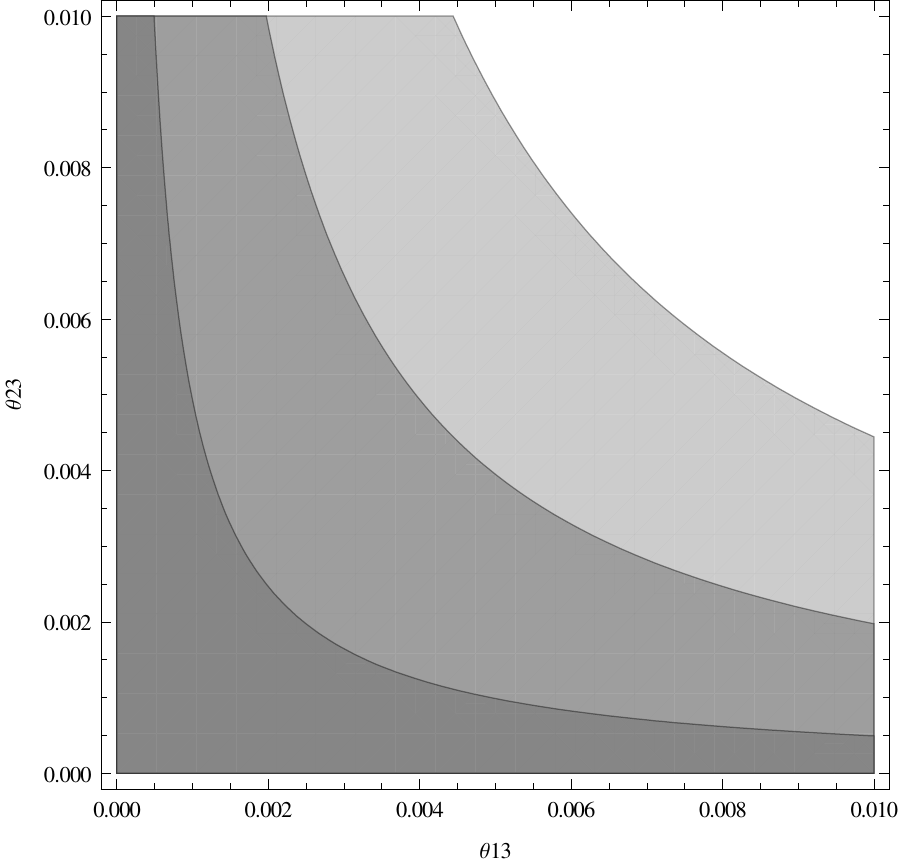}
\end{center}
\caption{Bounds coming from the $\mu\to e\gamma$ process in the different planes such that the third mixing angle is setting to zero for $m_{Z'}(1000, 2000, 3000 GeV)$.}
\label{3muegamma}
\end{figure}

In order to do the numerical analysis, we trace back the final parameters which are going
to be present in the decay widths and they are basically the mixing angles $\theta_{12}$, $\theta_{23}$,  $\theta_{13}$ and the $Z'$ gauge boson mass. There are also phases coming from the $V_l$
matrix but we have check that they are not important and we will neglect them. We are going to consider two cases depending which leptonic family is
assigned in the different representation of $SU(3)_L$: the first leptonic family in the different representation or the third leptonic family in the different representation. We should mention that
the option of the second leptonic family in a different representation is completely analogous to the case of the first family, so we do not present that case.

For the case of the first leptonic in a different representation, the rotation matrix in the charged leptonic sector is depending on  $\theta_{12}$, $\theta_{13}$ and the $Z'$ boson mass, assuming that
the phases involved are zero. Now, we can use the experimental bounds on the different LFV processes in order to get constraints on the mixing parameters and the $Z'$ boson mass. 
We have explored three different planes:
 $\theta_{13}$ vs $\theta_{12}$ with $\theta_{23}=0$, $\theta_{23}$ vs $\theta_{12}$ with $\theta_{13}=0$ and $\theta_{23}$ vs $\theta_{13}$ with $\theta_{12}=0$. Then for the process $\mu\to e\gamma$ considering $\theta_{23}=0$ with any small value $\theta_{13}<0.01$, bounds on $\theta_{12}$  are $\{0.2,1,2.2\} \times 10^{-3}$ for 
$M_{Z'}=\{1000,2000,3000 \}$ GeV respectively. And taking $\theta_{13}=0$ or $\theta_{12}=0$, that scenarios are plotted in
 \ref{1muegamma}. For the process $\tau\to e\gamma$, taking the parameter $\theta_{23}=0$ with small values of $\theta_{12}$, bounds on $\theta_{13}=\{0.2,1.7,3.8\}\times 10^{-3}$ for $Z'$ boson masses $M_{Z'}=\{1000,2000,3000 GeV\}$ are gotten. For the same process but taking  $\theta_{12}=0$ with small values of $\theta_{23}$ bounds on $\theta_{13}=\{0.4,1.8,4\}\times 10^{-3}$ are gotten for $M_{Z'}=\{1000,2000,3000 GeV\}$ respectively.

Using the experimental bounds on the process $\tau\to\mu\gamma$, we have gotten $\theta_{23}=\{1,3,7\}\times 10^{-3}$ taking
 $\theta_{13}=0$ or $\theta_{12}=0$ for $M_{Z'}=\{1000,2000,3000 GeV\}$. And using bounds on  $\mu \to eee$, we have gotten
$\theta_{12}=\{1,3.5,8\}\times 10^{-4}$ taking $\theta_{23}=0$ or  $\theta_{13}=0$ for the same values of $Z'$ boson mass. Finally, we have considered the processes $\tau \to lll$ with the parameter $\theta_{23}=0$ and any small value of $\theta_{12}$, bounds on $\theta_{13}$ are $\{0.2,0.8,1.8\}\times 10^{-2}$ for $M_{Z'}=\{1000,2000,3000 GeV\}$. But if we take $\theta_{13}=0$ for any small value of $\theta_{12}$, the bounds are on $\theta_{23}=\{0.5,1.8,3.8\}\times 10^{-2}$ for $M_{Z'}=\{1000,2000,3000 GeV\}$ respectively.


Now, we are going to consider the case when the third leptonic family is in a different representation of $SU(3)_L$, then the mixing matrix is
 depending on three mixing angles $\theta_{12}$, $\theta_{23}$ and $\theta_{13}$ and some phases that again we have taken equal to zero because they are not relevant.
Again we consider the same LFV processes used in the first case already mentioned. For the  $\mu\to e\gamma$ process, bounds in the plane  $\theta_{23}-\theta_{13}$ are obtained when $\theta_{12}=0$ is fixed, that plane is shown in figure
\ref{3muegamma}. We have also explored the other LFV processes but they do not put stringent constraints on the mixing angle parameters than the obtained with $\mu\to e\gamma$.

This work has been supported in part by UNAL-DIB grant 14844 and JAR acknowledge the finnancial support by DIB movilidad.

\end{document}